\documentclass[aps, twocolumn]{revtex4}

\usepackage{amsmath}
\usepackage{graphicx}
\usepackage[sort&compress]{natbib}

\newcommand{\ch}{\mathrm{ch}}
\newcommand{\sh}{\mathrm{sh}}
\renewcommand{\tanh}{\mathrm{\,th}}
\newcommand{\ctanh}{\mathrm{\,cth}}

\renewcommand{\Re}{\mathrm{Re}}
\renewcommand{\Im}{\mathrm{Im}}

\renewcommand{\section}[1]{

\textit{#1.}
}

\begin{document}

\title{Theory of helicity-sensitive terahertz radiation detection by field effect transistors}
\author{K. S. Romanov}
\author{M. I. Dyakonov}
\affiliation{Laboratoire Charles Coulomb, Universit\'e Montpellier II, CNRS, France}

\begin{abstract}
Within the two antenna model, we develop a theory of the recently observed helicity-sensitive detection of terahertz radiation by FETs. The effect arises because of the mixing of the ac signals produced in the channel by the two antennas. We calculate the helicity-dependent part of the photoresponse and its dependence on the antenna impedance, gate length, and gate voltage.
\end{abstract}

\maketitle

\textit{Introduction.}
Field effect transistors (FETs) can be used for efficient detection 
of terahertz radiation 
\cite{1996:dyakonov:shur,2001:dyakonov:shur,2010:dyakonov,2009:knap:dyakonov}. 
The standard model \cite{1996:dyakonov:shur} 
assumes that the radiation is coupled to the transistor by an 
effective antenna which generates an ac voltage predominantly
on one side of the transistor (e.g. between source and gate contacts).
Since the ac gate-to-channel voltage modulates both the electric field
and the electron concentration in the channel, the current density
will contain a dc component, which leads to a photoresponse in the form
of a dc source-drain voltage $\Delta U$ proportional to the
radiation intensity. Terahertz radiation detection and imaging
by FETs was demonstrated in many experimental works, see reviews in 
Refs.~\cite{2009:knap:dyakonov,2013:knap:dyakonov}.

For a number of applications it is essential to characterize 
the polarization of terahertz radiation. The strong sensitivity of FET 
detectors to linear polarization has been demonstrated
\cite{2008:sakowicz,2008:knap,2008:kim} and was shown to be
caused by the anisotropic sensitivity of
the effective antennas (formed by bonding wires and contact pads). 
However the problem of general terahertz polarimetry, including 
the determination of circular or elliptical polarization is still open.
The first step in this direction was made in the recent experiments of
Drexler et al \cite{2012:ganichev}, who discovered the strong sensitivity 
of the FET detector to the helicity of terahertz radiation, i.e. to the sign 
of the phase shift between the $x$ and $y$ components of the radiation field.

The qualitative explanation of this finding is based on the assumption 
(which was verified experimentally) that there are \emph{two} effective antennas 
connected to the source and the drain sides of the transistor. One of them is 
predominantly sensitive to the $x$-polarization and the other one~-- to 
$y$-polarization. For circularly polarized radiation, the ac voltages 
of these antennas are thus phase-shifted by $90^\circ$. To detect the helicity, 
the photoresponse must depend on this phase shift. This in turn is possible 
only because of the mixing of the ac signals produced in the channel 
by the two antennas. 

To understand the conditions for such mixing one should take into 
account that (i) the condition $\omega\tau\ll 1$ is normally satisfied, 
where $\omega$ is the radiation frequency and $\tau$ is the momentum 
relaxation time and (ii) under this condition the ac current injected 
at the source leaks to the gate through the distributed gate-to-channel
capacitance on a distance defined by the ac leakage length $l$ \cite{2010:dyakonov}. 
Thus the mixing of ac signals produced by the source and drain antennas
can occur only if the gate length $L$ is comparable to the leakage
length $l$. If $L\gg l$, the mixing becomes exponentially small.

This article presents a theoretical study of the helicity-sensitive detection 
of terahertz radiation by FETs within the two-antenna model \cite{2012:ganichev} explained above. 
We show that to describe the
required mixing of the two signals one has to abandon the usual
assumption that the antennas can be considered as ac voltage sources
(i. e. having zero ac impedance). Such mixing can occur
only when the antennas impedances are comparable to or greater than
the transistor impedance, and in particular when the antennas are
current sources (infinite impedance).  We calculate the FET photoresponse,
including its helicity-sensitive component, as a function of the gate
length $L$ and gate voltage $V_g$. This component has a maximum for certain
values of these parameters.

\begin{figure}[ht] 
\includegraphics[width=\columnwidth]{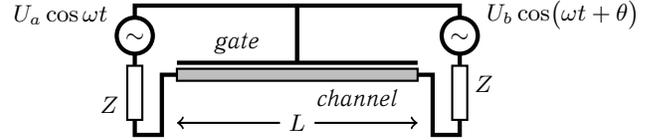}
\caption{Schematics of the high-frequency circuit of FET with source and drain antennas.} 
\end{figure} 
\textit{The problem}.
We consider a FET with antennas at both the source and the
drain sides, as presented in Fig.~1. The antennas are assumed
to generate ac voltages $U_a \cos\omega t$ and
$U_b \cos\bigl(\omega t + \theta\bigr)$ respectively, $\theta$ being the phase shift.
To simplify the calculations we assume that the antennas
have equal impedances at the radiation frequency $\omega$ denoted by $Z$.
We also assume that a dc voltage $V_g$ is applied between
the source and the gate, while the dc condition at the drain is open
circuit. We consider the case $\omega\tau \ll 1$. 

The basic equations are the continuity equation and the Ohm's law
\cite{2010:dyakonov}:
\begin{equation}
\label{Continuity_equation_and_Ohms_law}
\frac{\partial \rho}{\partial t} + \frac{\partial j}{\partial x} = 0,
\hspace{1cm}
j(x)=-\sigma \frac{\partial U}{\partial x},
\end{equation}
where $\rho$ and $j$ are the electron charge and current densities in
the channel, $\sigma=\rho\mu$ is the 2D channel conductivity, 
$\mu$ is the electron mobility, $U = V_{g} - V_{th}$ is the gate
voltage swing, $V_{th}$ is the threshold voltage.

We assume that the spatial variation of $U$ is large
compared to the gate-to-channel separation. Then the values
of $\rho$ and $\sigma$ are determined only by the local value of $U$
(the gradual channel approximation). For an open channel ($U>0$),
the plane capacitor formula is applicable:
\begin{equation}
\label{plain_capacitor_formula}
\rho=CU.
\end{equation}
Here $C$ is the gate-to-channel capacitance per unit
area. This relation is not valid below threshold ($U<0$) as well as in the vicinity
of the point $U=0$. To simplify the following calculations we will
derive our results assuming Eq.~(\ref{plain_capacitor_formula}) to
hold and the electron mobility $\mu$ to be constant. We will then 
present the modifications needed to account for the general dependences
$\rho(U)$ and $\sigma(U)$.

Combining Eqs.~(\ref{Continuity_equation_and_Ohms_law}) and
(\ref{plain_capacitor_formula}) we obtain the non-linear diffusion
equation
\begin{equation}
\label{general_equation_for_U}
\frac{\partial U}{\partial t}=\frac{\mu}{2}
    \frac{\partial^2 U^2}{\partial x^2}.
\end{equation}

We expand $U(x,t)=U_0 + U_1(x,t) + U_2(x,t)$ up to the second order 
of perturbations induced by radiation, i.e. $U_1$ is the
first order correction to $U_0$ and $U_2$ is the
second order correction. In the first order 
Eq.~(\ref{general_equation_for_U}) yields:
\begin{equation}
\label{linear_eq_u1}
\frac{\partial U_1}{\partial t}=
     \mu U_0 \frac{\partial^2 U_1}{\partial x^2}.
\end{equation}
The boundary conditions corresponding to Fig. 1 are:
\begin{equation}
\label{boundary_conditions}
\begin{array}{ll}
U_a \cos \omega t - j Z = U_1 & \mbox{at $x=0$},\\
U_b \cos (\omega t + \theta) + j Z = U_1 & \mbox{at $x=L$},
\end{array}
\end{equation}
where $Z$ is the antenna impedance.

From now on we will use dimensionless variables, assuming that
the voltages are in units of $U_0$, the coordinate $x$ and
the gate length $L$ are in units of the leakage length 
$l = \sqrt{2 \sigma/\omega C}=\sqrt{2 \mu U_0/\omega}$ \cite{2010:dyakonov},
and time $t$ is in units of $1/\omega$. In these units 
Eqs.~(\ref{linear_eq_u1}),~(\ref{boundary_conditions}) are
rewritten as:
\begin{eqnarray}
\label{unitless_eq_for_u1}
\frac{\partial U_1}{\partial t}=
     \frac{1}{2}\frac{\partial^2 U_1}{\partial x^2}, \\
\label{unitless_bc_for_u1}
\begin{array}{ll}
    U_a \cos t + \zeta \frac{\partial U_1}{\partial x} = U_1 & 
                \mbox{at $x=0$},\vspace{1mm}\\
    U_b \cos (t + \theta)-\zeta \frac{\partial U_1}{\partial x} = U_1 & 
                \mbox{at $x=L$},
\end{array}
\end{eqnarray}
where $\zeta = Z\sigma_0/l$ is the dimensionless antenna impedance,
$\sigma_0$ is the channel conductivity  in the absence of radiation.
In fact, $\zeta$ is the ratio of $Z$ and the resistance of
the rectifying part of the channel of length $l$.

To find the photoresponse voltage we need to consider the second
order equation following from Eq.~(\ref{general_equation_for_U}):
\begin{equation}
\label{linear_eq_u2}
\frac{\partial U_2}{\partial t}=\frac{1}{2}\left( 
    \frac{\partial^2 U_2}{\partial x^2} + 
    \frac{1}{2}\frac{\partial^2 U_1^2}{\partial x^2}
        \right).
\end{equation}
We average this equation over time and integrate twice over
$x$ \cite{2010:dyakonov}. The constant that appears after
the first integration is the dc current through FET which
should be zero due to the dc open circuit condition on the
drain side. Thus the photoresponse is given by
\begin{equation}
\label{general_DeltaU_exp}
\Delta U = \langle U_2(L) - U_2(0) \rangle = 
    \frac{1}{2}
        \Big[\langle U_1(0)^2 \rangle - \langle U_1(L)^2 \rangle \Big],
\end{equation}
where the angular brackets denote time-averaging.
If the impedance of antennas is negligible (i.e. $Z=0$),
then $U_1(x=0) = U_a \cos\omega t$ and 
$U_1(x=L) = U_b\cos(\omega t + \theta)$. For this case we obtain
the simple result:
\begin{equation}
\label{voltage_sources_response}
\Delta U = \frac{1}{4}(U_a^2 - U_b^2).
\end{equation}
Therefore, if the antennas are ac voltage sources, their signals
are not mixed, and consequently there is no sensitivity to the
phase shift $\theta$.

\textit{Solution.}
The general solution of equation Eq.~(\ref{unitless_eq_for_u1}) is
\begin{equation}
\label{general_solution}
U_1(x,t) = \frac{e^{i t}}{2} 
    \Big[A e^{(1+i)(x - L/2)} + B e^{(1+i)(L/2 - x)}\Big] + c.c.,
\end{equation}
where $c.c.$ means complex conjugated. Therefore
\begin{equation}
\label{averaged_u1_square}
\langle U_1(x,t)^2 \rangle = \frac{1}{2} 
    \Bigl|
        A e^{(1+i)(x-L/2)} + B e^{(1+i)(L/2-x)}
    \Bigr|^2.
\end{equation}
The boundary conditions Eq.~(\ref{unitless_bc_for_u1}) 
give the equations for the coefficients $A$ and $B$:
\begin{eqnarray}
\label{system_for_coeff}
\alpha A + \beta B = U_a,\nonumber\\
\beta A + \alpha B = U_b e^{i\theta},
\end{eqnarray}
with $\alpha = (1 - \eta) e^{-\lambda}$ and 
$\beta = (1 + \eta) e^{\lambda}$, $\lambda = (1+i)L/2$,
$\eta = (1+i)\zeta$. Using Eq.~(\ref{averaged_u1_square}) we can
now present the photoresponse voltage given by
Eq. (\ref{general_DeltaU_exp}) in the form
\begin{multline}
\Delta U = \frac{1}{4}\left\{
    \left|A e^{-\lambda} + B e^{\lambda} \right|^2 - 
    \left|A e^{\lambda} + B e^{-\lambda} \right|^2\right\} = \\
\Re\Bigl[
    \left(B - A\right) \left(A^* + B^*\right)
    \sh\lambda\; \ch\lambda^*
   \Bigr],
\end{multline}
where the star denotes complex conjugation. With the help
of Eq.~(\ref{system_for_coeff}) we obtain:
\begin{equation}
\label{general_delta_U}
\Delta U = 
\Re\Bigl[
    \frac{U_a - U_b e^{i\theta}}{\beta - \alpha}
    \left(
    \frac{U_a + U_b e^{i\theta}}{\alpha + \beta}
    \right)^*
    \sh\lambda\; \ch\lambda^*
   \Bigr].
\end{equation}
Finally, the expression for the photoresponse $\Delta U$ reads:
\begin{equation}
\label{delta_u_final}
\Delta U =\frac{1}{4} F_0 (U_a^2-U_b^2) + \frac{1}{2}F_1 U_a U_b \sin\theta,
\end{equation}
where the coefficients $F_0$ and $F_1$ are determined by
the dimensionless antenna impedance $\zeta$ and the dimensionless
gate length $L$, entering via the parameters $\eta$ and $\lambda$:
\begin{eqnarray}
\label{general_F0_F1}
F_0 = \Re(G), \hspace{1cm} F_1 = \Im(G),\nonumber\\
G = \frac{1}
         {(1+\eta \ctanh\lambda)(1+\eta \tanh\lambda)^*}.
\end{eqnarray}

We note that the dependence of $\Delta U$ on $U_a$, $U_b$,
and $\theta$ in Eq.~(\ref{delta_u_final}) naturally follows
from the symmetry of our problem resulting from the simplifying
assumption that the impedances of source and drain antennas
are equal. In this case, interchange of $U_a$ and $U_b$
together with the reversal of the phase shift $\theta$
should obviously change the sign of the photoresponse $\Delta U$. 

In particular, the interference term containing $U_aU_b$,
changes sign when $\theta$ is replaced by $-\theta$.
We recall that source and drain antennas are assumed
to predominantly respond to $x$- and $y$-polarizations respectively,
meaning that the contribution of the interference term
has opposite signs for right and left circular polarization.
If the impedances of the antennas were different,
$\sin\theta$ would be replaced by $\sin(\theta+\theta_0)$,
where $\theta_0$ is an additional intrinsic phase shift
related to the impedance difference.

Let us consider some limiting cases.

\textit{a}) Antennas are ac voltage sources, i.e. $Z=0$ and $\eta = 0$.
Then Eqs. (\ref{delta_u_final}) and (\ref{general_F0_F1}) give
$F_0=1$, $F_1=0$ and we retrieve the simple Eq.~(\ref{voltage_sources_response}).

\textit{b}) Antennas are ac current sources, i.e. $\zeta \gg 1$.
In this case it is convenient to introduce the antennas' ac
current amplitudes $j_a = U_a/|\eta|$ and $j_b = U_b/|\eta|$.
Using Eqs. (\ref{delta_u_final}) and (\ref{general_F0_F1}), we obtain:
\begin{equation}
\Delta U =\frac{1}{4} f_0 (j_a^2-j_b^2) + \frac{1}{2}f_1 j_a j_b \sin\theta,
\end{equation}
where
\begin{equation}
    f_0=\frac{\sh^2 L - \sin^2 L}{\sh^2 L + \sin^2 L},
\hspace{0.6 cm}
    f_1= \frac{2\sh L \sin L}{\sh^2 L + \sin^2 L}.
\end{equation}

\textit{c}) Long gate, $L\gg 1$. In this case the interference
between source and drain ac signals vanishes, and we obtain
the result similar to Eq. (\ref{voltage_sources_response}):
\begin{equation*}
\Delta U = \frac{1}{4}
     \frac{\left(U_a^2 - U_b^2\right)}{|1+\eta|^2},
\end{equation*}
which accounts for the finite antenna impedance.
\begin{figure}[ht]
\label{F1_vs_zeta}
\includegraphics[width=\columnwidth]{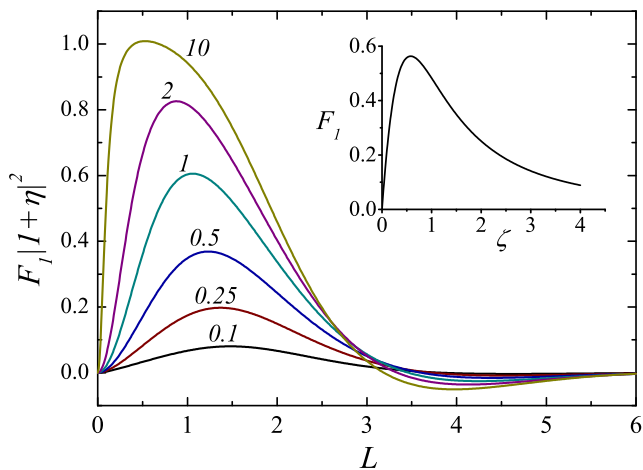}
\caption{The coefficient $F_1$ defining the helicity-sensitive part of
the photoresponse $\Delta U$ as a function of the gate length $L$
(in units of leakage length $l$) for different values of
the dimensionless antenna impedance $\zeta$, as indicated.
For convenience of presentation, $F_1$ is multiplied by
the factor $|1+\eta|^2=|1+(1+i)\zeta|^2$. Note the change of
sign around $L=\pi$. Inset: dependence of $F_1$ on $\zeta$ at $L=1$.}
\end{figure}

We are mostly interested in the helicity-sensitive part of
the photoresponse described by the coefficient $F_1$ in
the second term of Eq. (16). Fig. 2 presents $F_1$ as
a function of the gate length for different values of
the dimensionless antenna impedance $\zeta$. Maximum sensitivity
to radiation helicity occurs when the gate length is around
the leakage length $l$. As a function of $\zeta$ for a given $L$, 
the coefficient $F_1$ has a maximum around  $\zeta=1$, which corresponds to ac impedance matching, see the inset in Fig. 2.

\textit{Modifications for sub-threshold regime}. So far we considered detection
by an open channel ($U=V_g-V_{th}>0$) when Eq. 2 is valid. However experimentally the maximum of the photoresponse occurs \textit{below} threshold \cite{2009:knap:dyakonov, 2013:knap:dyakonov}. Fortunately, our theory can be easily extended to arbitrary gate voltages including sub-threshold values. 

Our solution of the basic Eq. 1 relied on the simple expressions $\rho=CU$ and $\sigma=\mu CU$ for the charge density and conductivity respectively, as well as on the implicit assumption that the mobility $\mu$ does not depend on $U$. This is a good approximation in the open channel regime ($U>0$). Generally, one has to consider $\rho$ and $\sigma$ as non-linear functions of $U$ not neccessarily proportional to each other.
Then the derivatives $\partial \rho / \partial t$ and $\partial \sigma / \partial x$, appearing in Eq. 1, should be expessed as 
\begin{equation}
 \frac{\partial\rho}{\partial t}=\frac{\partial\rho}{\partial U}\frac{\partial U}{\partial t}, \hspace{0.8cm}
 \frac{\partial\sigma}{\partial x}=\frac{\partial\sigma}{\partial U}\frac{\partial U}{\partial x}.
\end{equation}

For the simplest situation (one ac voltage source antenna exciting a long gate FET), this consideration leads to an important generalization of the expression $\Delta U = U_a^2/4U_0$ \cite{1996:dyakonov:shur}, which reads \cite{2012:lifshits, 2011:sakowicz}: 
\begin{equation}
 \Delta U=\frac{1}{4}\Big(\frac{1}{\sigma}\frac{\partial\sigma}{\partial U}\Big)_{U=U_0}U_a^2.
\end{equation}
This formula which replaces $1/U_0$ by $(\partial \ln \sigma /\partial U)_{U=U_0}$ is valid for arbitrary gate voltage, it also takes into account the possible dependence $\mu(U)$.

Using Eqs. 20, it can be shown that in our problem the generalisation to arbitrary dependences $\rho (U)$ and $\sigma (U)$ can be reduced to the simple change in units of length (leakage length $l$) and voltage (gate voltage swing $U_0$):  
\begin{equation}
 l\rightarrow \Big (\frac{\omega}{2\mu}\frac{1}{\rho}\frac{\partial\rho}{\partial U}\Big )^{-1/2}_{U=U_0}, \hspace{0.4cm} U_0\rightarrow \Big (\frac{1}{\sigma}\frac{\partial\sigma}{\partial U}\Big )^{-1}_{U=U_0}.
\end{equation}
With these modifications of units, the previous results in Eqs. (16, 17) remain valid.

\textit{Gate voltage dependence.} Now we can analyze the dependence of the helicity-sensitive part of the photoresponse on the gate voltage swing $U_0$. This dependence is governed by the interplay of several factors: (i)change of the ac impedance of the active parts of the transitor and corresponding variation of the parameter $\zeta$, (ii)change of the ratio $L/l$ due to the variation of the leakage length $l$, and (iii)variation of the voltage unit in Eq. 22.

Assuming the mobility $\mu$ to be constant, we will use for $\rho (U)$ the semi-empirical formula \cite{1996:shur}:
\begin{equation}
 \rho = CU^* \ln [1+\exp(U/U^*)], 
\end{equation}
where $U^*=\alpha kT/e$, $T$ is the temperature, and $\alpha$ is a phenomenological parameter on the order of 1. This formula reduces to Eq. 2 for positive $U\gg U^*$ and gives an exponential decrease of $\rho$ for negative $U$. 

As explained above, the generalized expression for the photoresponse is given by Eq. 16 provided the voltages $U_a$, $U_b$, and $\Delta U$ are measured in voltage units in Eq. 22. However this unit is inconvenient because it itself depends on $U_0$. For this reason we will switch to a constant voltage unit $U^*$.  Then Eq. 16 will have the same form, except that the coefficients $F_0$ and $F_1$, should be replaced by 
\begin{equation*}
\mathcal{F}_0=F_0 U^* \Big (\frac{1}{\sigma}\frac{\partial\sigma}{\partial U}\Big )_{U=U_0}, \hspace{0.2cm} \mathcal{F}_1=F_1 U^* \Big (\frac{1}{\sigma}\frac{\partial\sigma}{\partial U}\Big )_{U=U_0}.
\end{equation*}
We have performed numerical calculations of $\mathcal{F}_1$ using Eq. 23. Fig. 3 shows the dependence of $\mathcal{F}_1$ on the dimensionless gate voltage swing $U_0/U^*$ for several values of the antenna impedance. Since the dimensionless impedance $\zeta$ changes with gate voltage, we label the curves by the values of $\zeta$  at threshold ($U_0=0$). Also, the gate length $L$ was taken equal to the leakage length $l$ at threshold.

One can see that the helicity-dependent contribution to the photoresponse has a maximum around or below threshold. The decrease at the left side is due to the growth of the channel resistance. As a consequence, the dimensionless antenna impedance decreases approaching the limit $\zeta=0$ with no sensitivity to helicity, see the case \textit{a}) above. The decrease at the right side results from the drop of the ratio $L/l$ (see Fig. 2) caused by the increase of the leakage length $l$.

Consistent with our results, an increase of the helicity-dependent contribution to the FET photoresponse when moving towards negative gate bias was observed experimentally \cite{2012:ganichev}. However, the existing experimental data is not sufficient for a detailed comparison with our results.
\begin{figure}[ht]
\label{F1_vs_Vg}
\includegraphics[width=0.98\columnwidth]{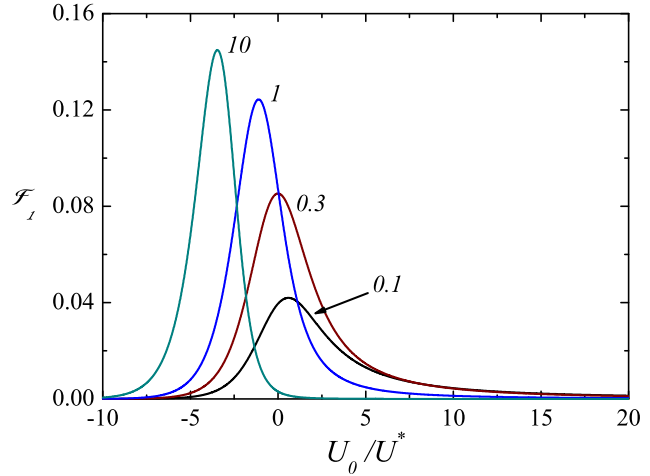}
\caption{The helicity-sensitive contribution $\mathcal{F}_1$ as a function of $U_0/U^*$ for different values of $\zeta$, as indicated, and $L=l$ (both $\zeta$ and $l$ taken at $U_0=0$).}
\end{figure}

In summary, we have presented a theory of helicity-dependent terahertz detection by FET based on the model with two antennas connected to the source and drain sides of the transistor and sensitive to orthogonal linear polarizations. We have found the dependence of the helicity-sensitive component of the photoresponse on the antenna impedance, gate length, and gate voltage.

\textit{Acknowledgement.} We thank Nina Dyakonova, Sergey Ganichev, and Maria Lifshits for interesting discussions.

\end{document}